\documentclass{aa}
\usepackage{xymtex}
\usepackage{newtxtext,newtxmath}
\usepackage{graphicx} 
\usepackage{subcaption}
\usepackage{amsmath} 
\usepackage{xcolor}
\usepackage{float}
\usepackage{caption}
\usepackage{placeins}
\usepackage{amsmath}
\usepackage{lipsum}
\usepackage{multicol}
\usepackage{soul}
\usepackage[T1]{fontenc}
\usepackage{comment}

\def\refjnl#1{{\rm#1}}
\def\apjl{\refjnl{ApJ}}                
\def\apjs{\refjnl{ApJS}}               
\def\ao{\refjnl{Appl.~Opt.}}           
\def\aap{\refjnl{A\&A}}                

\def\be{\begin{equation}} 
\def\ee{\end{equation}}

\def\msun{{\msun}}

\def\HI{\hbox{H~$\scriptstyle\rm I\ $}}

\def\gsim{\lower.5ex\hbox{\gtsima}} 
\def\lsim{\lower.5ex\hbox{\ltsima}} \def\gtsima{$\; \buildrel > \over 
\sim \;$} \def\ltsima{$\; \buildrel < \over \sim \;$} \def\prosima{$\; 
\buildrel \propto \over \sim \;$} \def\gsim{\lower.5ex\hbox{\gtsima}} 
\def\lsim{\lower.5ex\hbox{\ltsima}} 
\def\simgt{\lower.5ex\hbox{\gtsima}} 
\def\simlt{\lower.5ex\hbox{\ltsima}} 
\def\simpr{\lower.5ex\hbox{\prosima}}   
  
 \def\gtsima{$\; \buildrel > \over \sim \;$} 
\def\ltsima{$\; \buildrel < \over \sim \;$} 
\def\gsim{\lower.5ex\hbox{\gtsima}} 
\def\lsim{\lower.5ex\hbox{\ltsima}} 
\def\simgt{\lower.5ex\hbox{\gtsima}} 
\def\simlt{\lower.5ex\hbox{\ltsima}} 
\def\simpr{\lower.5ex\hbox{\prosima}}

\def\E3{{\cal E}_{\rm g}^{III}} 
\def\msun{\rm M_\odot}

\def\zsun{\rm Z_\odot}

\def\M*{M_*}

\def\Z*{Z_*}
\def\L*{L_*}
\def\muv{\rm M_{UV}}

\def\faccb{f_{\rm bh}^{\rm ac}}
\def\maccb{M_{bh}^{ac}}
\def\med{M_{\rm Edd}}
\def\fed{f_{\rm Edd}}
\def\mcritb{M_{\rm bh}^{\rm crit}}

\def\fescsf{f_{\rm esc}^{\rm sf}}
\def\fescbh{f_{\rm esc}^{\rm bh}}

\makeatletter
\renewcommand*\aa@pageof{, page \thepage{} of \pageref*{LastPage}}
\makeatother

\begin{document} 
   \title{UNCOVERing the contribution of black holes to reionization}

   \author{Pratika Dayal
          \inst{1}
          \and Marta Volonteri
          \inst{2}
          \and Jenny E. Greene
          \inst{3}
          \and Vasily Kokorev
          \inst{1}
          \and Andy D. Goulding
          \inst{3}
          \and Christina C. Williams
          \inst{4}
          \and Lukas J. Furtak
          \inst{5}
          \and Adi Zitrin
          \inst{5}
          \and Hakim Atek
          \inst{6}
          \and Rachel Bezanson
          \inst{7}
          \and Iryna Chemerynska
          \inst{6}
          \and Robert Feldmann
          \inst{8}
          \and Karl Glazebrook
          \inst{9}
          \and Ivo Labbe
          \inst{9}
          \and Themiya Nanayakkara
          \inst{9}
          \and Pascal A. Oesch
          \inst{10,11}
          \and John R. Weaver
          \inst{12}
          }

   \institute{Kapteyn Astronomical Institute, University of Groningen, PO Box 800, 9700 AV Groningen, The Netherlands\\
              \email{p.dayal@rug.nl}
              \and {Institut d'Astrophysique de Paris, CNRS, Sorbonne Universit\'e, 98bis Boulevard Arago, 75014, Paris, France}
              \and Department of Astrophysical Sciences, Princeton University, 4 Ivy Lane, Princeton, NJ 08544, USA
              \and NSF’s National Optical-Infrared Astronomy Research Laboratory, 950 North Cherry Avenue, Tucson, AZ 85719, USA
              \and Physics Department, Ben-Gurion University of the Negev, P.O. Box 653, Be’er-Sheva 84105, Israel
              \and Institut d'Astrophysique de Paris, CNRS, Sorbonne Universit\'e, 98bis Boulevard Arago, 75014, Paris, France
              \and Department of Physics \& Astronomy and PITT PACC, University of Pittsburgh, Pittsburgh, PA 15260, USA
              \and Department of Astrophysics, University of Zurich, Winterthurerstrasse 190, 8057 Zurich, Switzerland
              \and Centre for Astrophysics and Supercomputing, Swinburne University of Technology, PO Box 218, Hawthorn, VIC 3122, Australia
              \and Department of Astronomy, University of Geneva, Chemin Pegasi 51, 1290 Versoix, Switzerland
              \and Cosmic Dawn Center (DAWN), Niels Bohr Institute, University of Copenhagen, Jagtvej 128, K\o benhavn N, DK-2200, Denmark
              \and Department of Astronomy, University of Massachusetts Amherst, Amherst MA 01003, USA
             }


  \abstract
   {With its sensitivity in the rest-frame optical, the James Webb Space Telescope (JWST) has uncovered active galactic nuclei (AGN), comprising both intrinsically faint and heavily reddened sources, well into the first billion years of the Universe, at $z \sim 4-11$.    }
   {In this work, we revisit the AGN contribution to reionization given the high number densities associated with these objects. }
    {We use the {\sc delphi} semi-analytic model, base-lined against the latest high-redshift datasets from the JWST and the Atacama Large millimetre Array (ALMA) to model early star forming galaxies and AGN. We calculate the escape fractions of ionizing radiation from both star formation and AGN and include the impact of reionization feeback in suppressing the baryonic content of low-mass galaxies in ionized regions. This model is validated against the key observables for star forming galaxy, AGN and reionization.}
   {In our {\it fiducial} model, reionization reaches its mid-point at $z \sim 6.9$ and ends by $z \sim 5.9$. Low stellar mass ($M_*\lsim 10^9\msun$) star forming galaxies are found to be the key drivers of the reionization process, providing about $77\%$ of the total photon budget. Despite their high numbers, high accretion rates and higher escape fractions compared to star forming galaxies at $z \sim 5$, AGN only provide about $23\%$ of the total reionization budget which is dominated by black holes in high stellar mass systems (with $M_* \gsim 10^9\msun$). This is because AGN number densities become relevant only at $z \lsim 7$ - as a result, AGN contribute as much as galaxies as late as $z \sim 6.2$, when reionization is already in its end stages. Finally, we find that even contrasting models of the AGN ionizing photon escape fraction (increasing or decreasing with stellar mass) do not qualitatively change our results.}
   {}


   \maketitle

\section{Introduction}
\label{sec_intro}
The progress of reionization and its key sources remain outstanding questions in the field of physical cosmology. This is because the process of reionization depends on a number of poorly understood quantities including the abundance of galaxies in the first billion years, their star formation/black hole accretion rates that determine the production rate of ionizing photons, the fraction of such photons that can escape out of the galactic environment into the intergalactic medium (IGM) and the (spatially- and temporally-varying) clumping factor of the IGM that determines the recombination rate, to name a few \citep[for reviews see e.g.][]{dayal2018, robertson2022}. 

Consensus has been growing that low-mass galaxies are the most probable drivers of the reionization process, given their number densities that show a rising trend down to absolute magnitudes as faint as $\muv \sim -15$ \citep[e.g.][]{robertson2015, madau2017, dayal2020, trebitsch2022, atek2023, rinaldi2023}. Black holes seem to contribute at most a few tens of percent to the reionization process, mostly in its end stages at $z \lsim 5$ \citep[e.g.][]{onoue2017, dayal2020, trebitsch2023}. However, such results crucially depend on the number density of active galactic nuclei (AGN), especially in intermediate-mass halos (given that black hole accretion can produce 10 times as many ionizing photons as star formation per unit baryon) in addition to the escape fraction of ionizing photons from such sources \citep[e.g.][]{dayal2020}. Before the advent of the James Webb Space Telescope (JWST), a number of observations \citep[e.g.][]{giallongo2015, boutsia2018, giallongo2019, fujimoto2022} had already hinted at a larger than expected number density of faint AGN at $z>4$ although such results remained highly debated \citep{mcgreer2018, parsa2018, akiyama2018}. The proposal was that such number densities holding up to higher redshifts might imply a significant AGN contribution to the reionization process \citep[e.g.][]{grazian2018, mitra2018}. Overall, the relative importance of star forming galaxies and AGN to reionization has remained a key outstanding issue, mostly driven by a lack of statistically significant samples of black holes at $z \gsim 6$.

With its sensitivity, recent observations with the JWST have been crucial in shedding light on the AGN population at $z \gsim 4$: firstly, the detection of broadened Hydrogen Alpha ($H\alpha$) lines from deep JWST spectroscopy have been used to detect the ubiquitous presence of intrinsically faint AGN at $z \gsim 4$ \citep{harikane2023bh, maiolino2023, maiolino2023b}. Secondly, the JWST has yielded a new population of  intrinsically bright, reddened sources, ``little red dots' (LRDs), that exhibit extremely small sizes (of the order of a few hundred parsecs) and a characteristic ``V-shaped'' continuum which is red in the rest-frame optical but has a blue slope in the rest-frame ultra-violet \citep[UV;][]{labbe2023_nat, labbe2023, furtak2023_phot}. Spectroscopic follow-ups of these sources reveal clearly broadened Balmer lines which have been used to infer the presence of accreting black holes \citep[e.g.][]{furtak2023, kokorev2023, fujimoto2023_uncover, greene2023, matthee2023, kocevski2023, ubler2023, kokorev2024}. Whilst accounting for a few percent of the galaxy population at $z>5$, these relatively faint AGN seem to account for about $10-20\%$ of broad-line selected AGN at $z \sim 5-6$ \citep{harikane2023, maiolino2023b}. These combined observations have been used to infer black hole masses ranging between $10^{6.5-8}\msun$ \citep[]{harikane2023, maiolino2023, maiolino2023b, kocevski2023, furtak2023, larson2023, bogdan2023, goulding2023, kokorev2023, kokorev2024}. 

In this work, we revisit the black hole contribution to reionization in light of these newly-detected AGN well within the first billion years. In addition to new modes of black hole seeding and growth required to match to JWST observations, this model - termed {\sc delphi-dustbh} includes all of the relevant dust processes and has been base-lined against the latest dust observations from the Atacama Large millimetre Array (ALMA) surveys at $z \sim 5-7$ \citep{bethermin2020,bouwens2022}. This is particularly crucial given that observationally, the escape fractions of ionizing photons from star forming sources is closely linked to their dust enrichment \citep[e.g.][]{chisholm2022}. We start by describing the semi-analytic model used for this analysis in Sec. \ref{sec_model}, the escape fractions of ionizing photons from star forming galaxies and AGN in Sec. \ref{sec_fesc} and the reionization-feedback weighted emissivity in Sec. \ref{sec_reionization}. We then validate the model by showing its comparison to the ultra-violet luminosity function (UV LF) in Sec. \ref{sec_observables} before discussing the emissivity and escape fractions from galaxies and AGN in Sec. \ref{sec_ion}. We validate the model in Sec. \ref{sec_reio_obs}, detail the contribution of galaxies and AGN to reionization in Sec. \ref{sec_reio_sources} and explore the cumulative AGN contribution for two different models of black hole escape fractions in Sec. \ref{sec_fesc_bh} before concluding in Sec. \ref{sec_conclusions}.

We adopt a $\Lambda$CDM model with dark energy, dark matter and baryonic densities in units of the critical density as $\Omega_{\Lambda}= 0.673$, $\Omega_{m}= 0.315$ and $\Omega_{b}= 0.049$, respectively, a Hubble constant $H_0=100\, h\,{\rm km}\,{\rm s}^{-1}\,{\rm Mpc}^{-1}$ with $h=0.673$, spectral index $n=0.96$ and normalisation $\sigma_{8}=0.81$ \citep[][]{planck2020}. Throughout this work, we use a Salpeter initial mass function \citep[IMF;][]{salpeter1955} between $0.1-100 \msun$. Finally, we quote all quantities in comoving units, unless stated otherwise, and express all magnitudes in the standard AB system \citep{oke-gunn1983}.

\begin{table*}
\begin{center}
\caption{Free parameters (column 1), their symbols (column 2), the values used in the pre-JWST version of the {\sc delphi} model \citep[{\sc delphi2019,}][]{dayal2019} in column 3 and the values used in this work ({\sc delphi-dustbh}; column 4) that have been calibrated against datasets from both ALMA and the JWST. These differences and their physical implications are discussed in Sec. \ref{sec_model}. }
\begin{tabular}{|c|c|c|c|}
\hline
Parameter & Symbol & {\sc delphi2019} & {\sc delphi-dustbh} \\
\hline
Maximum star formation efficiency &  $f_*$ & $0.02$ & $0.1$ \\
Fraction of SNII energy coupling to gas & $f_{\rm w}$ & $0.1$ & $0.075$ \\
BH seed mass & - &   $150\msun$ & $10^{3-5}\msun$\\
Radiative efficiency of black hole accretion & $\epsilon_{\rm r}$ & $0.1$ & $0.1$\\
Fraction of AGN energy coupling to gas &  $f_{\rm bh}^{\rm w}$ & $0.003$ & $0.001$ \\
Fraction of gas mass AGN can accrete & $f_{\rm bh}^{\rm ac} (M_h < {\rm M_h^{crit}})$ &  $5.5 \times 10^{-4}$ & $5.5 \times 10^{-4}$ \\
Fraction of gas mass AGN can accrete & $f_{\rm bh}^{\rm ac} (M_h > {\rm M_h^{crit}})$ &  $5.5 \times 10^{-4}$ & 0.1 \\
Fraction of Eddington rate for BH accretion & $f_{\rm Edd} (M_h < {\rm M_h^{crit}})$ & $7.5 \times 10^{-5}$ & $10^{-4}$ \\
Fraction of Eddington rate for BH accretion & $f_{\rm Edd} (M_h \geq {\rm M_h^{crit}})$ & $1$ & $1$\\
Escape fraction of \HI ionizing photons from star formation & $\fescsf$ &  $0.02 [(1+z)/7]^{2.8}$ & $\fescsf ={\rm fn}(\beta)^{\dag}$ \\
Escape fraction of \HI ionizing photons from AGN & $f_{\rm esc}^{\rm bh}$ & \citet{ueda2014} & $\fescbh = {\rm fn}(f_{\rm Edd})^{\ddag}$ \\
Stellar population synthesis model & - &  {\sc Starburst99} & {\sc Starburst99} \\
Reionization (UVB) feedback & - & No & Yes\\
\hline
\end{tabular}
\tablefoot{$^{\dag}$ As detailed in Sec. \ref{sec_fesc}, the value of $\fescsf$ is determined using observed relations \citep{chisholm2022} that link $\fescsf$ to the UV spectral slope ($\beta$). \\
$^{\ddag}$ Further, $\fescbh$ is inferred from the Eddington accretion fraction ($f_{\rm Edd})$) based on the results of zoom-in hydrodynamic simulations \citep{trebitsch2019}. }
\label{table1}
\end{center}
\end{table*}

\section{The Theoretical model}
\label{sec_model}
This work is based on using the {\sc Delphi} ({\bf D}ark Matter and the {\bf e}mergence of ga{\bf l}axies in the e{\bf p}oc{\bf h} of re{\bf i}onization) semi-analytic model for galaxy formation that uses a binary merger tree approach to jointly track the build-up of dark matter halos, their baryonic components (gas, stellar, dust and metal masses) and black holes. This version, that tracks the seeding and assembly of black holes driven by recent JWST observations, and the dust enrichment of early sources motivated by ALMA results, is termed {\sc delphi-dustbh}. It follows the assembly of dark matter halos between $\log(M_h/ \msun)=8-14$ from $z \sim 40$ down to $z = 4.5$ in time-steps of 30 Myrs. At any time-step, the available gas mass (from both mergers and accretion) can form stars with an ``effective" star formation efficiency which is the minimum between the efficiency that produces enough type II Supernova (SNII) energy to eject the remainder of the gas and an upper maximum threshold ($f_*$). The two free parameters concerning star formation are $f_* \sim 10\%$ and the fraction of SNII energy that can couple to gas ($f_{\rm w} \sim 7.5\%$). Their values are obtained by simultaneously matching to the faint and bright-ends of the Lyman Break Galaxy (LBG) UV LF at $z \sim 5-9$, and to the dust-stellar mass relations obtained from the latest ALMA observations \citep{bethermin2020,bouwens2022}. The (about 5 times higher) value of $f_*$ here compared to our previous model {\sc delphi2019} \citep{dayal2019} is driven by dust attenuation which plays a specially crucial role for high-mass systems. In this case, we require a higher star formation efficiency to match to the observed bright-end of the UV LF \citep[for details see][]{dayal2022,mauerhofer2023}. Further, every newly formed stellar population (i.e. newly formed stellar mass within a given time-step) is assumed to have an age of 2 Myrs. This age is used in conjunction with its metallicity, that is tracked by our model, to calculate the ionizing photon production rate using the {\sc starburst99} \citep{leitherer1999} stellar population synthesis code.

Further, while the details of black hole physics (growth from both accretion and instantaneous mergers and the associated feedback and a radiative efficiency of $\epsilon_r = 10\%$) remain the same as in our previous works \citep{dayal2019, dayal2020}, we now summarise the key new ingredients required to match to the AGN populations being observed by the JWST: 

\begin{enumerate}
\item Explaining the black hole masses inferred for a number of observed early black holes requires black hole seeds heavier than those from Population III stars which have masses of the order of $\sim 150\msun$ \citep[as discussed in e.g.][]{kokorev2023, furtak2023, maiolino2023b, natarajan2023} as used in our previous works \citep{dayal2019}. We therefore seed the starting halos of any merger tree at $z \gsim 13$ with heavy seeds of masses (randomly chosen) between $10^{3-5} \msun$. Widespread seeds with masses $\sim 10^{3} \msun$ can form in dense, massive stellar clusters through a number of pathways \citep[for a review see Sec. 2.3.1][]{amaro-seoane2023} including dynamical interactions \citep[e.g.][]{devecchi2009}, the runaway merger of stellar mass black holes \citep[e.g][]{belczynski2002} or the growth of stellar mass black holes in conjunction with mergers \citep[e.g.][]{leigh2013}. They can grow to larger masses within the clusters \citep{Alexander2014}. Additionally, seeds with masses at birth up to $10^{5} \msun$ can form via supermassive star formation, sometimes referred to as direct collapse black holes, although this mechanism produces a lower number of viable seeds \citep{dayal2019}.

\item Our model includes a ``critical" halo mass for efficient black hole accretion with a value that evolves with redshift as $\mcritb(z) = 10^{11.25}[\Omega_m(1+z)^3 +\Omega_\Lambda ]^{-0.125}$ on which we include a scatter of 0.5 dex, motivated by the results of cosmological simulations \citep[e.g.][]{bower2017}. In order to explain the number density of JWST-detected AGN we modify some of the parameters. Black holes are allowed a gas accretion rate of $\maccb(z) = min[\epsilon_r \faccb  M_{\rm g}^{\rm sf}, \fed \med]$ where $M_{\rm g}^{\rm sf}$ is the gas mass left after star formation and its associated SNII feedback, $\fed$ is the Eddington fraction and $\med$ is the Eddington accretion rate. Allowing very weak AGN feedback ($0.01\%$ of black hole feedback coupling to the gas) we require values of $\faccb = 0.1 ~ (5 \times 10^{-4})$ and $\fed = 1.0 ~ (10^{-4})$ for halos above (below) the critical mass (we allow 0.5 dex of scatter on all of these quantities). That is, black holes in high-mass (low-mass) halos can accrete the minimum between 10\% ($0.05\%$) of the available gas mass and $100\% ~(0.01\%)$ of the Eddington fraction. High-gas accretion rates in halos above the critical halo mass are crucial to simultaneously match to AGN observables including the bolometric and UV LFs, and black hole mass functions.  

\item We include the impact of dust attenuation on the luminosity and ionizing photon escape fractions for both star formation and AGN, as detailed in Sec. \ref{sec_fesc}. 

\item Finally, we account for a coupling of reionization feedback and galaxy formation in order to determine the role of star formation and AGN in the reionization process as detailed in Sec. \ref{sec_reionization}.

\end{enumerate}

\subsection{Dust and the escape fraction of ionizing photons}
\label{sec_fesc}
The escape fraction of ionizing photons from star forming galaxies and AGN remain a key outstanding issue in the field of reionization \citep[e.g. Sec. 7.1,][]{dayal2018}. In this work, we use simple phenomenological prescriptions to obtain the escape fraction from star formation ($\fescsf$) and AGN ($\fescbh$). For star forming galaxies \citep[as in our previous work,][]{trebitsch2022}, we use the results from the Low-z Lyman Continuum survey \citep[LzLCS][]{flury2022} that targets 66 $z \sim 0.3$ Lyman continuum emitters which can be treated as analogues of early sources. For this sample, $\fescsf$ is closely linked to the UV spectral slope ($\beta$) between 1300-1800\AA~ as characterised in \citet{chisholm2022}:
\begin{equation}
\fescsf = (1.3 \pm 0.6) \times 10^{-4} \times 10^{(-1.22 \pm 0.1)\beta}
\end{equation}
Further, high-$z$ observations \citep[e.g.][]{bouwens2014} show a correlation between the $\beta$ slope and the observed (dust attenuated) UV magnitude such that \citep[see e.g.][]{trebitsch2022}
\begin{equation}
\beta = -1.993 - 0.071(z - 6) - 0.125(\muv + 19.5).
\end{equation}
The above relations imply an increase in $\fescsf$ with a decrease in dust attenuation i.e. with bluer $\beta$ slopes. We calculate the dust attenuated magnitude for star forming galaxies using the model detailed in \citet{dayal2022}. Essentially, we use the latest metal yields from \citet{kobayashi2020} and calculate the dust mass including the key processes of dust formation, destruction, ejection, astration and grain growth in the interstellar medium (ISM); we assume a grain growth timescale of 30Myr timescale. This has been shown to yield results in accord with the latest ALMA observations for $z \gsim 5$ galaxies \citep{dayal2022, mauerhofer2023}. This dust is assumed to be perfectly mixed with the gas in order to calculate the optical depth to UV photons to obtain the dust-attenuated UV magnitude. 

As for the dust attenuation of black holes, we use results from zoom-in simulations that estimate the column density of gas around black holes in high-redshift galaxies \citep{trebitsch2019}. These simulations find that gas density in the central region of galaxies ($\sim 40$ pc), which dominates the total column density from the AGN to the galaxy edge, is modulated by inflows and outflows, leading to a scaling of the form:
\begin{equation}
\frac{N_{H}}{10^{22}{\rm cm^{-2}}} = \bigg(\frac{f_{\rm Edd}}{0.01}\bigg)^{0.5}.
\end{equation}
Note that this does not include an explicit ``torus''. However, given the high gas content in high-$z$ galaxies the ISM column density is a significant, if not dominant, component \citep[see][]{Circosta2019}. Assuming a 0.5 dex of scatter on this relation, the inferred column density can be used to calculate the UV optical depth as $\tau_{bh}(\lambda) = \int n_d \sigma_d(\lambda) dl$ where $n_d$ is the number density of dust grains and $\sigma_d$ is the dust interaction cross-section per unit hydrogen atom using the fits from \citet{gnedin2008} based on the Small Magellanic Cloud (SMC) model. Further, given that we only model the total gas, dust and metal masses in our model, we assume the dust number density to be $n_d = n_g (Z/\zsun)$ where $n_g$ is the number density of gas particles and $Z$ is the gas phase metallicity normalised to the mean SMC metallicity of $\zsun=0.005$. We then assume the same escape fraction for UV and ionizing photons from AGN such that $\fescbh = e^{-\tau_{bh}}$. 

\begin{figure*}
\begin{center}
\center{\includegraphics[scale=1.01]{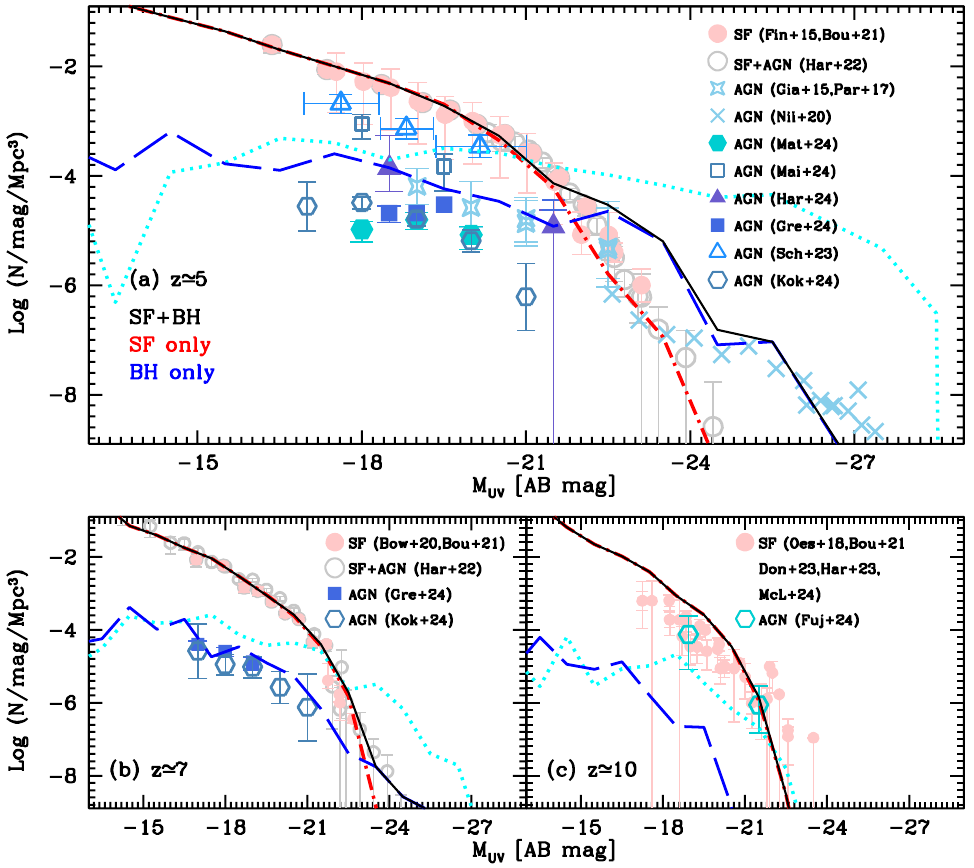}} 
\caption{The rest-frame UV LF at $z \sim 5$, $7$ and $10$, as marked. In each panel, the long-dashed, dot-dashed and solid lines show the dust obscured UV LF for AGN, star formation and the ``total" luminosity, as marked; dotted cyan lines show the intrinsic (i.e. dust-unattenuated) AGN UV LF. In all panels, points show observational results as marked: at $z \sim 5$ (panel a), for star forming galaxies \citep[solid circles,][]{finkelstein2015,bouwens2021}, galaxies+AGN \citep[empty circles,][]{harikane2021}, AGN \citep[empty stars and crosses,][]{giallongo2015,parsa2018, niida2020} and the new JWST AGN results from \citet[filled hexagons]{matthee2023}, \citet[empty squares]{maiolino2023}, \citet[filled triangles]{harikane2023}, \citet[empty triangles]{scholtz2023}, \citet[filled squares]{greene2023} and \citet[empty hexagons]{kokorev2024}. At $z \sim 7$ (panel b), we show the results for star forming galaxies \citep[solid circles,][]{bowler2020,bouwens2021}, galaxies+AGN \citep[empty circles,][]{harikane2021} and the new JWST AGN results from \citet[filled squares]{greene2023} and \citet[empty hexagons]{kokorev2024}. At $z \sim 10$ (panel c), we show the UV LF inferred for star forming galaxies \citep[solid circles,][]{oesch2018,bouwens2021,donnan2022,harikane2023b,mcleod2024}. At this redshift, the empty hexagons show AGN candidates (UHZ1 and GN-z11) with the number densities calculated as detailed in \citet{fujimoto2023_uncover}. }
\label{fig_uvlf}
\end{center}
\end{figure*}

\begin{figure*}
\begin{center}
\center{\includegraphics[scale=1.1]{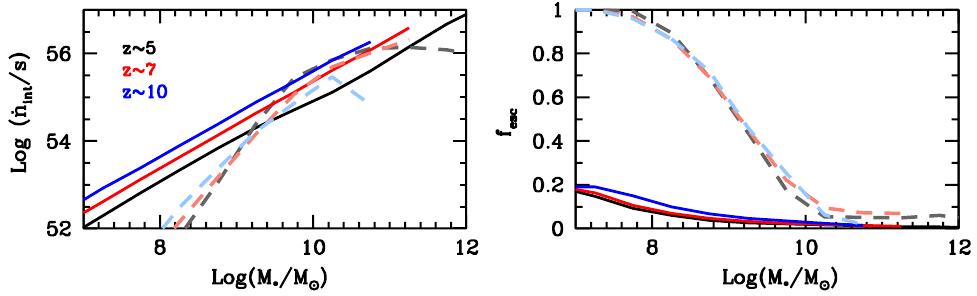}} 
\caption{As a function of the stellar mass, we show the intrinsic production rate of ionizing photons ({\it left panel}) and their escape fractions ({\it right panel}) at $z \sim 5-10$, as marked. In each panel, the solid and dashed lines show the results for star formation and black holes, respectively. }
\label{fig_ionfesc_fnms}
\end{center}
\end{figure*}
\subsection{The progress of reionization and its impact on galaxy formation}
\label{sec_reionization}
We can now calculate the volume filling fraction of ionized hydrogen ($Q_{\rm II}$) to track the progress of reionization as \citep{shapiro1987, madau1999,dayal2020}:
\begin{equation}
\label{filfracz}
\frac{dQ_{\rm II}}{dz} = \frac{dn_{\rm ion}} {dz}\frac{1}{n_{\rm H}} - \frac{Q_{\rm II}}{t_{\rm rec}} \frac{dt}{dz}.
\end{equation}
Here, $\dot n_{ion}= dn_{\rm ion}/dz$ is the total escaping rate of ionizing photons (from both star formation and AGN), $n_{\rm H}$ is the comoving hydrogen number density and  $t_{\rm rec}$ is the recombination timescale for ionized hydrogen - we use an IGM clumping factor of $C =  1+ 43 \, z^{-1.71}$ \citep{pawlik2009} to calculate $t_{\rm rec}$ \citep[for complete details see][]{dayal2020}.

The build-up of the heating UV background (UVB) from reionization can suppress the baryonic content of low-mass halos in ionized regions \citep[e.g.][]{gnedin2000,sobacchi2013b, hutter2021}. We account for this process by running a ``maximal UVB" scenario where we completely suppress the gas mass (and the associated star formation and black hole accretion rate) for halos with virial velocities below $50~{\rm Km~s^{-1}}$; this corresponds to halo masses of $10^{9.7} ~(10^{9.3})\msun$ at $z \sim 5~ (10)$. The total emerging ionizing emissivity for reionization is then obtained by weighing over the UV-suppressed contribution from low-mass halos in ionized regions ($Q_{\rm II}$) as well as that from un-suppressed sources in neutral regions ($Q_{\rm I} = 1-Q_{\rm II}$) such that \citep{choudhury2019}
\begin{eqnarray}
\label{eq_nion}
\nonumber
  \dot n_{\rm ion}(z) & = & \fescsf [Q_{\rm II}(z) \dot n_{\rm int,II}^{\rm sf}(z)  + Q_{\rm I}(z) \dot n_{\rm int,I}^{\rm sf}(z) ] \\ 
  & & + f_{\rm esc}^{\rm bh}[Q_{\rm II}(z) \dot n_{\rm int,II}^{\rm bh}(z) + Q_{\rm I}(z) \dot n_{\rm int,I}^{\rm bh}(z)],
\end{eqnarray}
where the first and second terms on the right hand-side account for the contribution of star forming galaxies and AGN, respectively. As might be expected, while un-suppressed halos dominate at the start of the reionization process, the importance of the UV-suppressed term increases as reionization proceeds. This is the {\it fiducial} model that is used throughout the paper. 

\section{Early black holes and the reionization process in light of JWST data}
\label{sec_results}
We now validate our model against the UV LF for both star forming galaxies and AGN in Sec. \ref{sec_observables} before discussing their associated ionizing emissivities and escape fractions in Sec. \ref{sec_ion}. We validate the model against key reionization observables in Sec. \ref{sec_reio_obs} and discuss the key reionization sources in Sec. \ref{sec_reio_sources} before ending by showing the AGN contribution for two contrasting models of $\fescbh$ in Sec. \ref{sec_fesc_bh}.

\subsection{The UV LF for star forming galaxies and AGN}
\label{sec_observables}
We start by discussing the dust-attenuated UV LF from star forming galaxies, AGN and the total UV luminosity (summed over both these components) as shown in Fig. \ref{fig_uvlf}. As noted, our choice of the model parameters for star formation (maximal threshold star formation efficiency and the fraction of SN energy coupling to gas) and dust \citep[consistent with our previous works, e.g.][]{dayal2022} are tuned to match to the observed UV LF for LBGs at $z \sim 5-9$ and for AGN at $z \sim 5-7$ in addition to the stellar and black hole mass functions and the AGN bolometric LF. Therefore, by construction, the model yields a dust-attenuated LBG UV LF that is in good agreement with the observations (including those from the JWST), within error bars, for $\muv \sim -16$ to $-23.5$ at $z \sim 5-7$. This also naturally results in derived quantities such as the UV luminosity density being in accord with observations \citep[for complete details see e.g.][]{dayal2022, mauerhofer2023}. 

As seen from the same figure, our model for AGN dust attenuation yields an evolving UV LF that is in good accord with JWST observations of both LRDs and faint AGN (with $\muv \sim -18$ to $-21.5$) at $z \sim 5$ \citep{matthee2023, maiolino2023, maiolino2023b, harikane2023} and the UV LF at $z \sim 5-7$ from \citet{greene2023}; our results are also in accord with earlier AGN UV LF estimates \citep[e.g.][]{giallongo2015, parsa2018}. However, the model over-predicts the AGN UV LF between $\muv \sim -23$ to $-24$ as compared to the observations from \citet{niida2020}. This is possibly due to the combination of a simple model for dust attenuation, a mass-independent black hole accretion rate and an up-scattering of low-mass black holes into these high luminosity bins due to the scatter of 0.5 dex assumed on all of the model parameters. A further possibility is that pre-JWST observations may underestimate the bright end of the AGN UV LF.  The UV luminosity functions are generally based on spectroscopically confirmed sources from photometric selected samples, and thus may suffer from some incompleteness. It is also important to keep in mind that in some cases, the AGN signatures are detected in the rest-frame optical, and we have incomplete knowledge of the origin of the UV emission which may therefore have a significant contribution from the stellar component. In the case of the little red dots we still do not know the origin of the UV \citep[e.g.][]{labbe2023, furtak2023_phot, greene2023}. Finally, for the X-ray selected sources without spectroscopic follow-up, we do not know the unobscured/obscured fractions \citep[e.g.][] {giallongo2015, parsa2018}.

We find star formation to dominate the faint-to-intermediate end ($\muv \gsim -21.5$) of the deconstructed UV LF at $z \sim 5$ with AGN dominating at brighter luminosities \citep[see also][]{ono2017, piana2022}. We find UV faint AGN to contribute $\sim 0.4-15\%$ to the total number density at intermediate luminosities ($\muv \sim -18$ to $-21$) at this redshift. By $z \sim 7$, AGN dominate the UV LF at much brighter luminosities corresponding to $\muv \sim -23$ - this is naturally expected as both the number densities and masses of massive black holes build up with time. As seen, we find that the intrinsic (dust-unattenuated) AGN UV LF severely over-estimates the bright end of the AGN UV LF; an increasing impact of dust attenuation with decreasing redshift is crucial to match the AGN UV LF to observations.

The stellar component dominates the UV at all magnitudes at $z \sim 10$ where massive black holes have not yet been able to assemble in significant numbers. Indeed at $z \sim 10$, the AGN UV LF predicted by the model is a factor 15-40 lower in the case the number densities of AGN inferred from UHZ1 and GN-z11 correctly represent the AGN population at this redshift \citep{fujimoto2023_uncover}. We however caution that the nature of these two sources is poorly known. Indeed, it is highly plausible that the observed UV luminosity has a significant contribution from star formation rather than being powered by AGN accretion alone. Within error bars, these observed number densities are in accord with the ``intrinsic" (i.e. dust-unattenuated) AGN UV LF from our model. Accounting for these objects and assuming an upper limit of $\fescbh=1$ would yield the maximum AGN contribution to reionization. The results of this model are discussed in Sec. \ref{sec_reio_sources} where we show that such a model can be ruled out since it severely overshoots the observed emissivity constraints at $z \lsim 6$.

\subsection{The ionizing emissivity and its escape fraction from star formation and black holes}
\label{sec_ion}
We now discuss the intrinsic production rate of ionizing photons as a function of stellar mass from both star forming galaxies ($\dot n_{\rm int}^{\rm sf}$) and AGN ($\dot n_{\rm int}^{\rm bh}$) at $z \sim 5-10$, as shown in (panel a of) Fig. \ref{fig_ionfesc_fnms}. We find that $\dot n_{\rm int}^{\rm sf}$ scales with $M_*$ at all $z \sim 5-10$ since in our model, both the stellar mass assembly and the instantaneous star formation rate are closely tied to the underlying halo potential. The mean relation for star formation can be described by $\dot n_{\rm int}^{\rm sf} = 0.9 M_* + \delta$ where $\delta = (45.9,46.3,46.6)$ at $z \sim (5, 7, 10)$, respectively. The amplitude of the relation increases with redshift because galaxies of a given stellar mass are hosted in halos of similar masses at all $z \sim 5-10$. However, halos of a given mass correspond to deeper potentials with increasing redshift which result in an increase in the star formation rate and therefore an increase in the production rate of ionizing photons. The behaviour of AGN is more mass-dependent \citep[see e.g.][]{dayal2020}: for example, at $z \sim 5$, $\dot n_{\rm int}^{\rm bh}$ essentially scales with the stellar mass for $M_* \sim 10^{8-10}\msun$ galaxies, thereafter flattening out to $M_* \sim 10^{12}\msun$ - this is driven by the fact that although high mass galaxies have an availability of gas, the black hole can only accrete the minimum between 10\% of the gas mass and 100\% of the Eddington limit as discussed in Sec. \ref{sec_model}. At $z \sim 5$, $\dot n_{\rm int}^{\rm bh}$ exceeds $\dot n_{\rm int}^{\rm sf}$ for $M_* \gsim 10^{9.2-11.4}\msun$ galaxies. Indeed, for $M_* \sim 10^{10}\msun$ systems, black holes produce about 5 times as many ionizing photons ($\dot n_{\rm int}^{\rm bh} \sim 10^{55.8} {\rm s^{-1}}$) as star formation. The same qualitative behaviour persists at $z \sim 7$ where black holes contribute roughly equally as star formation to the ionizing photon production rate for $M_* \sim 10^{10-11}\msun$ galaxies which is of the order of $\dot n_{\rm int}^{\rm bh} \sim 10^{55-56} {\rm s^{-1}}$. At $z \sim 10$, however, massive enough black holes have not had time to assemble, as a result of which the stellar component dominates the ionizing photon production rate for all stellar masses. 

\begin{figure*}
\begin{center}
\center{\includegraphics[scale=1.05]{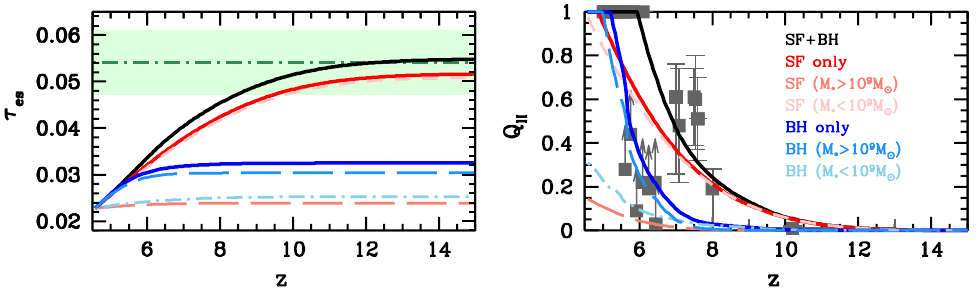}} 
\caption{The redshift evolution of the electron scattering optical depth ({\it left panel}) and the volume filling fraction of ionizied hydrogen ({\it right panel}). We show the volume filling fraction weighted values from both star forming galaxies and AGN as well as the contribution deconstructed into galaxies with stellar masses above and below $10^9\msun$, as marked. In the left panel, the dot-dashed horizontal line and the shaded region show the electron scattering optical depth and its associated error bars from \citet{planck2020}. In the right panel, points show compilations of $Q_{\rm II}$ results from a number of works \citep{fan2006, davies2018, yang2020b, jung2020, lu2020, bosman2022, gaikwad2023, nakane2023}.}
\label{fig_tau_qii}
\end{center}
\end{figure*}

We then discuss the trend of the escape fraction of ionizing photons from star formation and AGN as shown in (panel b) of the same figure. Given its direct dependence on the dust mass that increases with $M_*$, $\fescsf$ too decreases with an increase in the stellar mass. For example, at $z \sim 5$ $\fescsf$ decreases from about 16\% for $M_*\sim 10^7 \msun$ to $\sim 0.6\%$ for $M_*\sim 10^{12} \msun$. For a given stellar mass, the slight decrease in the dust attenuation with increasing redshift results in a mild increase of $\fescsf$ with $z$ - for example, $\fescsf$ increases from 2.5\% to 5\% for $M_* \sim 10^9 \msun$ between $z \sim 5$ and $10$. As for AGN, black holes in galaxies with $M_* \lsim 10^{8.5}\msun$ show $\fescbh\gsim 80\%$ given their low Eddington accretion rates. We find $\fescbh$ to show a steep drop for more massive systems due to an increase in both the Eddington accretion rate and metallicity which drive up the dust optical depth. For example, by $M_* \sim 10^{10}\msun$, $\fescbh$ drops to about 15\% at $z \sim 5-7$. We caution that this simple phenomenological model assumes spherical symmetry in calculating the escape fraction of ionizing photons from both star formation and AGN - in principle, however, there might be (dusty) dust-free lines of sight that would result in much (lower) higher escape fractions.  

\subsection{Validating the model against reionization observables}
\label{sec_reio_obs}
We now validate the model results against the two key reionization observables - the electron scattering optical depth ($\tau_{es}$) and $Q_{II}$ - whose results are shown in Fig. \ref{fig_tau_qii}. The model agreement with the early star forming galaxy population and $\fescsf$ values that scale inversely with both stellar mass and redshift result in a $\tau_{es}$ value in good agreement with the central value inferred from \citet{planck2020} data as seen from (the left panel of) Fig. \ref{fig_tau_qii}; this result was expected based on our previous works that have used the {\sc delphi} model for reionization, \citep{dayal2020, trebitsch2023}. Given their larger number densities and higher $\fescsf$ values, we find low-mass ($M_* \lsim 10^9\msun$) star forming galaxies to be the key drivers of the reionization process, with higher mass galaxies having a negligible contribution. We also find that black holes in high-mass galaxies ($M_* \gsim 10^9\msun$) dominate the AGN contribution to reionization.

\begin{figure*}
\begin{center}
\center{\includegraphics[scale=0.85]{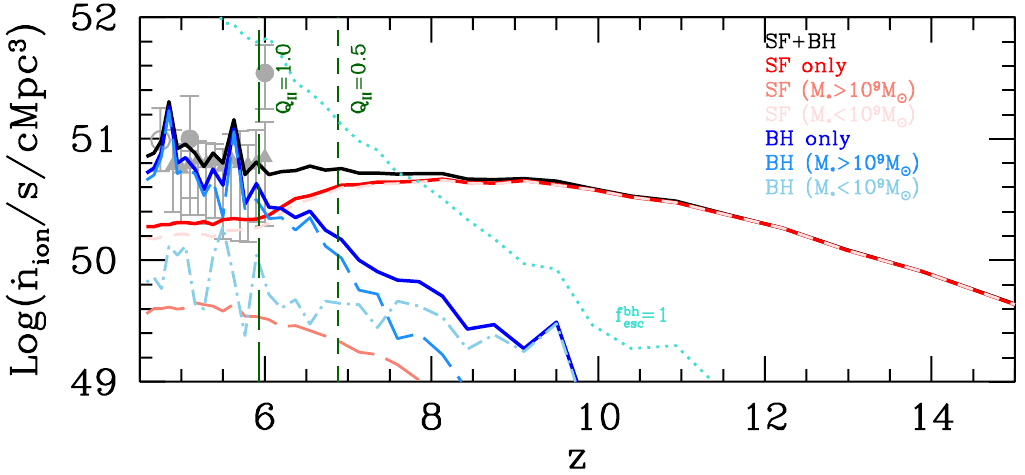}} 
\caption{The redshift evolution of the ``escaping" rate of ionizing photons from star forming galaxies and AGN weighted over the volume filling fraction to account for the effects of reionization feedback (Eqn. 5). The dotted line shows the ``maximal" AGN contribution assuming $\fescbh = 1$. We also show the total contribution deconstructed into sources with stellar masses above and below $10^9\msun$, as marked. Points show the observational results from \citet[empty circles]{becker2013}, \citet[filled circles]{becker2021} and \citet[filled triangles]{gaikwad2023}. The vertical short-dashed (long-dashed) line shows the redshift at which the IGM is half (fully) ionized i.e. $Q_{\rm II}=0.5$ ($1$), as marked.}
\label{fig_nion_weighted}
\end{center}
\end{figure*}

As seen from (the right panel of) the same figure, the contribution from both star formation and AGN result in a redshift evolution of $Q_{\rm II}$ that is in good agreement with a number of observational results within error bars \citep[e.g.][]{fan2006, davies2018, yang2020, jung2020, lu2020, bosman2022, gaikwad2023, nakane2023}; readers are also referred to \citep{fontanot2023} for a tabulated compilation. In this {\it fiducial} model, reionization reaches its mid-point at $z \sim 6.9$ and is over by $z_{\rm reio} \sim 5.9$. Low-mass ($M_* \lsim 10^9\msun$) star forming galaxies are the key drivers of reionization for the bulk of its history (discussed further in Sec. \ref{sec_reio_sources}). Indeed, while a scenario ignoring AGN shows no sensible effect on the progress of reionization down to $z \sim 7$, the end of the process is delayed to $z \sim 5$ (i.e. by about 200 Myrs). In an AGN-only scenario, on the other hand, the redshift evolution of $Q_{\rm II}$ is naturally delayed, with the mid-point of reionization being reached at $z \sim 6$. Driven by an increase in both the number densities and masses of black holes at $z \lsim 6$, reionization proceeds at an accelerated rate in this scenario and is over by $z \sim 5.2$, driven mostly by AGN in high-mass ($M_* \gsim 10^9\msun$) galaxies. An AGN-only scenario can be ruled out because it generates too low a value of $\tau_{es} = 0.032$ (as compared to the observed value of $0.054$) and is in tension with a number of inferred constraints on $Q_{\rm II}$ at $z \gsim 7$, as seen from the same figure.

We note that reionization ends at $z \sim 5.9$ in our {\it fiducial} model, while recent results on the Lyman$\alpha$ transmission in Quasar spectra seem to indicate reionization-related fluctuations (in the UVB, residual neutral hydrogen fraction, and/or IGM temperature)  to persist in the IGM down to $z \sim 5.3$, signalling a late end to the reionization process \citep{bosman2022}. This discrepancy could arise from a number of assumptions in our work that could have a significant bearing on the end of reionization as now detailed. Firstly, as detailed in Sec. \ref{sec_fesc}, we assign values of $\fescsf$ and $\fescbh$ based on the dust enrichment of the galaxy and the Eddington fraction, respectively. However, the escape of ionizing radiation is a highly stochastic process that depends on the line-of-sight \citep[e.g.][]{paardekooper2015, trebitsch2017} and is strongly modulated over timescales of Myrs due to both Supernova and radiative feedback \citep[e.g.][]{rosdahl2022}. Secondly, we assume complete suppression of the gas mass for halos in ionized regions below a fixed virial velocity of $50~{\rm Km~s^{-1}}$. Suppressing the gas content in more (less) massive halos by about 0.5 dex would shift the end stages of reionziation by $\delta z \sim 0.5$ as we have shown in a previous work \citep{trebitsch2022}. However this would impact both star formation and AGN growth, which would not significantly impact our results on the relative contributions of these sources to the reionization process. Finally, given the nature of the model, we can only track the volume filling fractions of neutral/ionized hydrogen rather than the reionization-related fluctuations measured by observations. We aim at fully implementing our semi-analytic model into a  hydrodynamic framework to be able to track these fluctuations at the end stages of reionization in future works.  

\subsection{The key sources of reionization}
\label{sec_reio_sources}
We now discuss the relative contribution of star forming galaxies and AGN to the reionization process accounting for its feedback (Eqn. 5) as shown in Fig. \ref{fig_nion_weighted}. As noted above, in our {\it fiducial} model, the mid-point of reionization occurs at $z \sim 6.9$ with reionization being over by $z \sim 5.9$. Star formation in low-mass galaxies (with $M_* \lsim 10^9\msun$) is the key driver of the reionization process, providing $>80\%$ of ionizing photons at any redshift between $z \sim 7.1-20$. As reionization proceeds and the volume filling fraction exceeds 50\% at $z \sim 6.9$, the gas masses and star formation rates of such sources are increasingly suppressed by UV feedback. This results in the emissivity from star formation turning over at $z \lsim 7$ and dropping from providing $75\%$ of ionizing photons at $z \sim 7$ to $33\%$ by the end of reionization at $z \sim 5.9$. With their lower number densities, star formation in high-mass galaxies (with $M_* \gsim 10^9\msun$) has a negligible contribution ($\lsim 5\%$) at any redshift between $z \sim 6-20$. 

As an increasing number of massive black holes assemble through efficient accretion, the AGN contribution to reionization increases with cosmic time from about $26\%$ at $z \sim 7$ to $66\%$ by $z \sim 5.9$. This contribution is originally driven by putative black holes in low-mass sources ($M_* \lsim 10^9\msun$) at $z \gsim 7.3$. However, as a result of their larger accretion rates (and hence production rate of ionizing photons) and the UV-suppression of low-mass systems, black holes in high-mass galaxies ($M_* \gsim 10^9\msun$) start dominating the AGN contribution to reionization at $z \lsim 7.3$ and make up $\sim 70\%$ of the total AGN budget by $z \sim 6$. However, despite their high numbers, high accretion rates and higher escape fractions (as compared to star formation), AGN only contribute as much as star forming galaxies in terms of the ionizing budget by $z \sim 6.2$ when reionization is already 80\% complete and therefore in its end stages. 

In terms of the cumulative photon budget in the {\it fiducial} model, the qualitative picture is very similar. Dominating the photon contribution for most of the reionization history, star formation in low-mass galaxies ($M_* \lsim 10^9\msun$) provides more than two-thirds ($\sim 77\%$) of the cumulative photon budget; star formation in higher-mass halos play a negligible role, providing $<3\%$ of the total reionization photon budget. As a result of their rising number densities at $z \lsim 7$, AGN provide a fourth (about $23\%$) of the reionization budget - this is mostly driven by AGN in high-mass halos ($M_* \gsim 10^9\msun$) which make up about $16.5\%$ of the budget with AGN in lower mass halos providing the final $\sim 6.5\%$ of reionization photons. Our result of star formation in low-mass galaxies being the key reionization driver \citep[e.g.][]{robertson2015, atek2023}, with AGN playing a role only towards the very end stages is in excellent accord with previous works \citep[e.g.][]{chardin2017, onoue2017, dayal2020, trebitsch2023}.

Finally, we show a ``maximal" AGN contribution case in which AGN are assumed to be completely dust-free at $z \gsim 7$ i.e. $\fescbh = 1$. This model is driven by the tentatively high amplitude of the AGN UV LF inferred at $z \sim 10$, which is in accord with a dust-free AGN scenario, as shown in Sec. \ref{sec_observables}. As seen from Fig. \ref{fig_nion_weighted}, in this case, AGN and galaxies contribute equally to the ionizing photon output by $z \sim 7.5$. However, given the steeply rising AGN contribution reionization finishes as early as $z \sim 7$ in this model. By this point, AGN contribute 36\% to the cumulative photon budget. However, as pointed out in our previous works \citep[e.g.]{dayal2020}, we reiterate that this case can be ruled out since it exceeds the emissivity values observed at $z \lsim 6$ \citep[by e.g.][]{becker2013, becker2021, gaikwad2023} in addition to over-estimating the observed AGN UV LF at $z \sim 5,7$ as shown in Sec. \ref{sec_observables}.

\subsection{The impact of the escape fraction of ionizing photons from AGN}
\label{sec_fesc_bh}

In order to study the impact of $\fescbh$ on our results, we study a contrasting model where we assume that the unobscured fraction of AGN is a proxy for $\fescbh$ \citep[see][for details and extensive comparison of different approaches on  $\fescbh$]{dayal2020}. We use the results from \citet{merloni2014} who find $\fescbh$ to scale with the intrinsic X-ray luminosity in the 2-10 keV range in a redshift-independent manner such that
\begin{equation}
\fescbh = 1-0.56 + \frac{1}{\pi} {\rm arctan} \bigg( \frac{43.89-{\rm log} L_X}{0.46}\bigg).
\end{equation}
This model results in a scenario where $\fescbh \propto M_*$ such that $\fescbh$ increases from $10\%$ to $95\%$ as $M_*$ increases from $10^{9.5}$ to $10^{11}\msun$; this is in contrast to the {\it fiducial} model where $\fescbh \propto M_*^{-1}$ and has a value of $\fescbh \sim 40\% ~ (5\%)$ for $M_* \sim 10^{9.5}~(10^{11})\msun$. 

\begin{figure}
\begin{center}
\center{\includegraphics[scale=0.5]{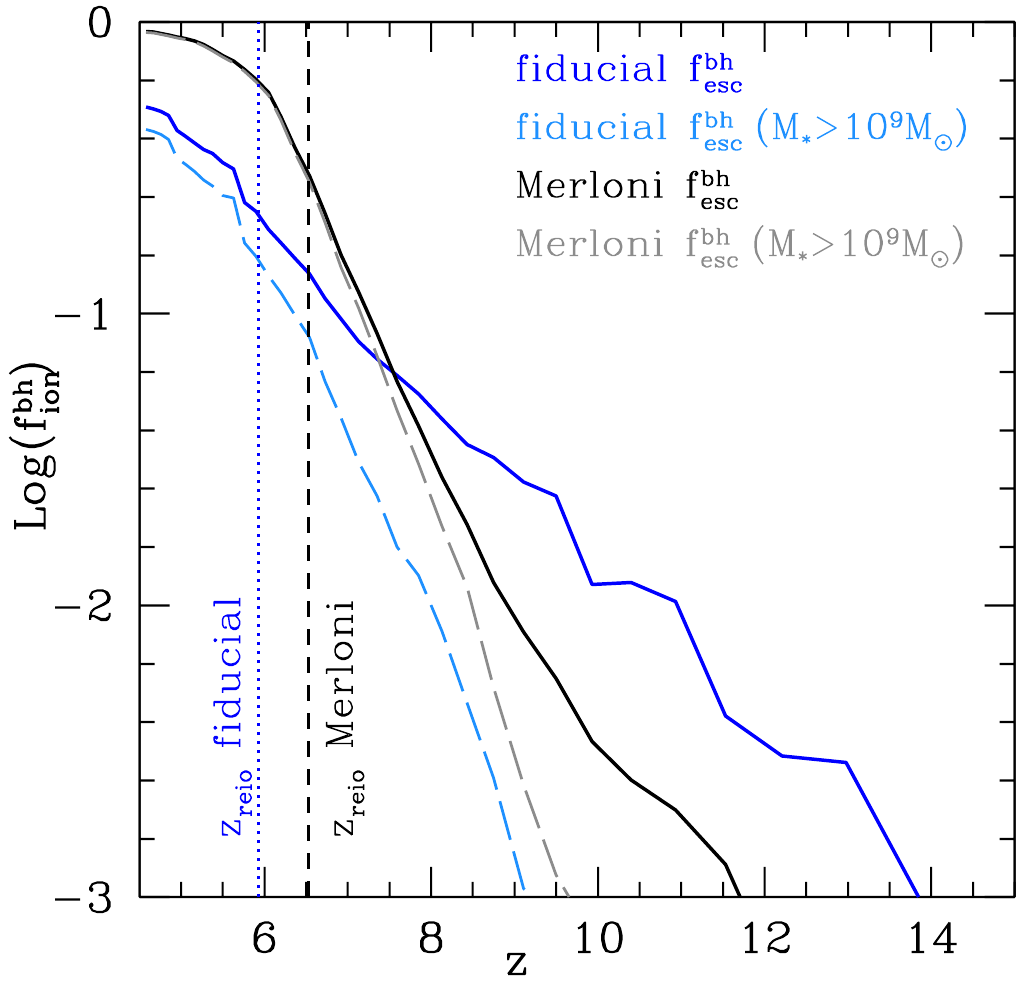}} 
\caption{The cumulative fraction of ionizing photons contributed by AGN as a function of redshift ($f_{\rm ion}^{\rm bh}$) as well as the contribution from AGN in high-mass ($M_* \gsim 10^9 \msun$) galaxies, as marked. We show results for two contrasting models of $\fescbh$: the first is the {\it fiducial} model where $\fescbh \propto M_*^{-1} $ and the second is the ``Merloni" \citep{merloni2014} model where $\fescbh \propto M_*$. We also show the redshifts ($z_{\rm reio}$) at which reionization finishes in both of these models.}
\label{fig_fracnion}
\end{center}
\end{figure}

We show the cumulative AGN contribution to reionization for both of these $\fescbh$ models in Fig. \ref{fig_fracnion}. Qualitatively, our results remain unchanged in that even in the ``Merloni" $\fescbh$ scenario, AGN contribute less than a third ($\sim 30\%$) of the photons to reionization by the end of the process at $z_{\rm reio} \sim 6.5$; this is slightly larger than the 22\% photons provided by AGN in the {\it fiducial} model by $z_{\rm reio} = 5.9$. A key difference between these models is that as a result of $\fescbh$ increasing with $M_*$, in the ``Merloni" model black holes in high-mass galaxies ($M_* \gsim 10^9\msun$) always make up the dominant AGN contribution. This model also shows a steeper slope of AGN contribution to reionization with decreasing redshift as increasingly massive black holes assemble and start contributing to reionization through a larger escaping output of ionizing photons.

\section{Conclusions and discussion}
\label{sec_conclusions}
With its sensitivity in the rest-frame optical, the JWST has yielded an unprecedented sample of black holes in the first billion years, comprising both intrinsically faint and reddened sources. Together, these objects could contribute as much as $15\%$ to the observed (i.e. dust attenuated) UV LF at intermediate luminosities ($\muv \sim -18$ to $-21$) at $z \sim 5$. We use the {\sc Delphi} semi-analytic model \citep{dayal2019}, tuned to the latest galaxy and AGN observables at $z \sim 5-9$, to revisit the black hole contribution to reionization in light of these results. We include heavy seeding mechanisms and allow high accretion rates to reproduce the observables for AGN at $z \sim 5-7$. In addition to a detailed model for the dust enrichment of early galaxies \citep[introduced in][]{dayal2020}, we include a phenomenological model for the dust attenuation of black holes. Finally, we use low-redshift analogues of Lyman continuum emitters to link the escape fraction of ionizing photons from star formation ($\fescsf$) to the dust-attenuated UV magnitude; we assume the escape fraction of ionizing photons from black holes ($\fescbh$) to be the same as the escape fraction of UV photons in the {\it fiducial} model. With this model, which matches the key reionization constraints (the electron scattering optical depth and the emissivity) and includes roughly 0.5 dex of scatter on all free parameters, our key findings are:

\begin{itemize}
\item While star forming galaxies dominate the faint-to-intermediate end of the UV LF ($\muv \gsim -21$), AGN determine the behaviour at the bright end at $z \sim 5$. However, the importance of AGN decreases with increasing redshift such that they are sub-dominant at all magnitudes by $z \sim 10$.

\item The intrinsic production rate of ionizing photons from star formation scale roughly linearly with $M_*$ as $\dot n_{\rm int}^{\rm sf} = 0.9 M_* +\delta$ where $\delta = (45.9, 46.6)$ at $z \sim (5,10)$. The production rate of ionizing photons from AGN is mass-and redshift-dependent: while AGN dominate over star-formation for $M_* \sim 10^{9.2-11.4}\msun$ galaxies at $z \sim 5$, their contribution becomes sub-dominant at all stellar masses by $z \sim 10$.

\item The value of $\fescsf$ decreases with $M_*$ as galaxies become more dust enriched. For example, at $z \sim 5$, $\fescsf$ drops from about $16\%$ to $<1\%$ as $M_*$ increases from $10^7$ to $10^{12}\msun$. On the other hand $\fescbh>80\%$ for $m_* \lsim 10^{8.5}\msun$ and shows a steep drop with increasing masses at all $z \sim 5-10$.

\item In the {\it fiducial} model, reionization reaches its mid-point at $z \sim 6.9$ and is over by $z \sim 5.9$. Low-mass ($M_* \lsim 10^9\msun$) star forming galaxies are the key drivers of the reionization process, providing $>80\%$ of ionizing photons at any redshift between $z \sim 7.1-20$; higher mass galaxies provide less than $5\%$ of ionizing photons at any redshift.  

\item As an increasing number of black holes start accreting efficiently with decreasing redshift, the AGN contribution rises from about $26\%$ at the mid-point of reionization ($z \sim 7$) to dominating the instantaneous budget ($\sim 66\%$) by the end-stages at $z \sim 5.9$. This is mostly driven by AGN in high-mass halos ($M_* \gsim 10^9\msun$).

\item Despite their high numbers, high accretion rates and higher escape fractions, AGN only contribute as much as star forming galaxies in terms of the ionizing budget by $z \sim 6.2$ when reionization is already in its end stages. 

\item Overall, low-mass star forming galaxies ($M_* \lsim 10^9\msun$) provide $\sim 77\%$ of the cumulative photon budget; star formation in higher mass halos play a negligible role, providing $<3\%$ of the total reionization photon budget. AGN provide about $23\%$ of the total reionization budget, mostly driven by AGN in high-mass halos ($M_* \gsim 10^9\msun$) with AGN in low-mass halos providing the final $\sim 6.5\%$ of reionization photons. 

\item A model with $\fescbh =1$ can be ruled out since it severely exceeds the observed emissivity constraints at $z \lsim 6$.

\item Finally we find that, provided $\fescbh<1$, even contrasting models of $\fescbh$ where $\fescbh \propto M_*$ or $\fescbh \propto M_*^{-1}$ do not affect our results significantly, with AGN providing less than a third of the total photon budget needed for reionization.

\end{itemize}

Our result of low-mass galaxies dominating the reionization process, with AGN affecting it in its end-stages are in good agreement with a number of previous works \citep{robertson2015, atek2023,chardin2017, onoue2017, dayal2020, trebitsch2023}. A key caveat, however, persists with regards to the exact nature and number densities of faint AGN. The origin of the UV emission is unclear in the case of the LRDs, and in Fig.~\ref{fig_uvlf} their observed UV luminosity is attributed solely to black hole accretion. However, for faint AGN, the UV luminosity could trace both the star forming and AGN components (Maiolino private communication). Detailed spectroscopic analyses will be crucial in shedding light on the nature and numbers of such objects. Additionally, combining the results from multiple surveys will be crucial in getting a handle on the cosmic variance associated with their number densities. We also caution that we have used an extremely simple geometrical model for the escape fraction model used for both star forming galaxies and AGN. While we assume spherical symmetry, the complex ISM might have a number of sight-lines that could allow a lower/higher escape fraction of both UV photons and ionizing radiation; this line-of-sight escape remains a key outstanding issue, solving which would benefit tremendously from spectroscopy of such objects. 

Over the next decade, we look towards cross-correlations of spectroscopically confirmed galaxies/AGN and tomographic 21cm data from (e.g.) the SKA. These will be crucial to shed light on the progress and key sources of reionization through the distribution and redshift evolution of ionized bubble sizes and their clustering \citep[e.g.][]{hutter2019}.

\begin{acknowledgements}
   PD acknowledge support from the NWO grant 016.VIDI.189.162 (``ODIN") and warmly thanks the European Commission's and University of Groningen's CO-FUND Rosalind Franklin program. J.E.G. acknowledges support from NSF/AAG grant 1007094, and also support from NSF/AAG grant 1007052.
   VK acknowledges funding from the Dutch Research Council (NWO) through the award of the Vici Grant VI.C.212.036. The research of CCW is supported by NOIRLab, which is managed by the Association of Universities for Research in Astronomy (AURA) under a cooperative agreement with the National Science Foundation. AZ acknowledges support by Grant No. 2020750 from the United States-Israel Binational Science Foundation (BSF) and Grant No. 2109066 from the United States National Science Foundation (NSF); by the Ministry of Science \& Technology, Israel; and by the Israel Science Foundation Grant No. 864/23. HA and IC acknowledge support from CNES, focused on the JWST mission, and the Programme National Cosmology and Galaxies (PNCG) of CNRS/INSU with INP and IN2P3, co-funded by CEA and CNES. KG and TN acknowledge support from Australian Research Council Laureate Fellowship FL180100060. This work has received funding from the Swiss State Secretariat for Education, Research and Innovation (SERI) under contract number MB22.00072, as well as from the Swiss National Science Foundation (SNSF) through project grant 200020\_207349. The Cosmic Dawn Center (DAWN) is funded by the Danish National Research Foundation under grant DNRF140. We also thank the referee for their insightful comments.

\end{acknowledgements}



\bibliographystyle{aa} 
\bibliography{mybib}

\end{document}